\documentclass[aps,pre,twocolumn,float]{revtex4}
\usepackage{amsmath,bm,epsfig}
\usepackage{amssymb}

\usepackage{graphicx}
\DeclareMathAlphabet{\mathitbf}{OML}{cmm}{b}{it}

\newcommand{\sFrac}[2]{{\textstyle\frac{#1}{#2}}}

\def\Fbox#1{\vskip1ex\hbox to 8.5cm{\hfil\fboxsep0.3cm\fbox{%
  \parbox{8.0cm}{#1}}\hfil}\vskip1ex\noindent}  %%  {TEXT} in BOX

\newcommand{\B}[1]{{\bm{#1}}}%% Bold Roman & Greek Lower & Upper Case
\let \= \equiv \let\*\cdot \let\~\widetilde \let\^\widehat \let\-\overline

\begin{document}
\title{Do Athermal Amorphous Solids Exist?}
\author{H.G.E. Hentschel$^{1,2}$,  Smarajit Karmakar$^1$, Edan Lerner$^1$ and Itamar Procaccia$^1$}
\affiliation{$^1$Dept. of Chemical Physics, The Weizmann Institute of
Sceince, Rehovot 76100, Israel
\\$^2$Dept of Physics, Emory University, Atlanta, GA 303322, }

\begin{abstract}
We study the elastic theory of amorphous solids made of particles with finite range interactions in the thermodynamic limit. For the elastic theory to exist one requires all the elastic coefficients, linear and nonlinear, to attain a finite thermodynamic limit. We show that for such systems the existence of non-affine mechanical responses results in anomalous fluctuations of all the nonlinear coefficients of the elastic theory. While the shear modulus exists, the first nonlinear coefficient $B_2$ has anomalous fluctuations and the second nonlinear coefficient $B_3$ and all the higher order coefficients (which are non-zero by symmetry) diverge in the thermodynamic limit.  These results put a question mark on the existence of elasticity (or solidity) of amorphous solids at finite strains, even at zero temperature. We discuss the physical meaning of these results and propose that in these systems elasticity can never be decoupled from plasticity: the nonlinear response must be very substantially plastic.
\end{abstract}
\maketitle
%%%%%%%%%%%%%%%%%%%%%%%%%%%%%%%%%%
\section{Introduction}
\label{Introduction}

Amorphous solids are ubiquitous in nature and in technology, spanning well known materials from obsidian to metallic glasses.
They are typically obtained by fast cooling a liquid such that it glides over the melting point where normally
it can form a crystalline solid via a standard first-order phase transition. Rather, the fluid becomes super-cooled and
experiences that so-called glass transition where its relaxation time becomes ever longer upon decreasing the temperature
\cite{09Cav}. When $T\to 0$ we have typically a frozen amorphous structure which appears to react elastically
to small strains. In this paper we consider only amorphous solids that are made of $N$ point particles in $d$
dimensions, in a volume $V$, which interact with each other via finite range interactions. The subject of our investigation is how
athermal amorphous solids react to external strains.

The potential energy in our strained amorphous solids can be written as  $U(\{\B r_i(\gamma )\}, \gamma)$,
where $\left\{\B r_i\right\}_{i=1}^N$ are the positions of the particles and $\gamma$ is the applied strain.
In this paper we shall treat the strain as a scalar, all equations can be written in full tensorial
form if required (cf. \cite{10KLPa}), but for our purposes a single scalar strain $\gamma$ will be sufficient.

For normal crystalline solids there is a rich literature spanning decades and centuries of
studies of their elastic properties. It is well known that for small external strains $\gamma$ normal solids react
elastically to create an opposed stress $\sigma$. For very small external
strains the stress is linear in the strain, with the coefficient of proportionality being the shear modulus $\mu$. Choosing
an equilibrium reference state at $\gamma=\gamma_0$ the stress resulting from a small strain thus reads
\begin{equation}
\sigma = \mu (\gamma-\gamma_0) \ .
\end{equation}
For higher strains the elastic response becomes nonlinear, and we denote, for the purpose of the discussion below,
the nonlinear coefficients as $B_n$ with $n\ge 2$ (that is to say, by convention $B_1\equiv \mu$). We thus write for
large strains
\begin{equation}
\sigma(\gamma) = \mu (\gamma-\gamma_0)+ B_2(\gamma-\gamma_0)^2+B_3 (\gamma-\gamma_0)^3 +\cdots \ . \label{sigofgam}
\end{equation}
For perfect monatomic crystalline solids the elastic theory is entirely transparent since both the linear and the nonlinear coefficients in this expansion can be computed directly from the change in system energy as a function of the strain,
\begin{equation}
B_n\to B^{\rm Born}_n=\frac{1}{V}\frac{\partial^{n+1} U}{ \partial \gamma^{n+1}} \ ,
\end{equation}
(the so-called Born term, and see below for details). When the crystalline solid becomes riddled with defects,
the nature of the elastic theory becomes much more obscure, since the Born term does not tell the whole story.
In amorphous solids deviations from the Born approximation are inevitable. Any strain is bound to bring about,
in addition to an affine transformation in the position of the particles, also non-affine transformations that are
necessary to bring back the system to a mechanical equilibrium in which the net force $\B f_i$ on every particle is zero.
Nevertheless, The fundamental assumption that is equivalent to the assertion that a piece of material is a solid,
is that before any plastic deformation takes place, the potential can be expanded in powers of the strain to derive
the various expressions for the stress, shear modulus and higher nonlinear elastic coefficients.

In the following we consider deformations via parameterized
transformations on the particle coordinates $\B J(\gamma) = \B I + \gamma\B h$ where $\B h$ determines the
imposed deformation \cite{footnote},
\begin{equation}\label{transformation}
\B r_i \to \B J\cdot \B r_i + \B u_i \ ,
\end{equation}
where the non-affine coordinates $\B u_i$ are additional displacements
that guarantee that the mechanical equilibrium constraint is fulfilled.
Total derivatives with respect to strain in the athermal limit
should thus satisfy the zero-forces constraint \cite{10KLPa,89Lut,04ML}:
\begin{equation}\label{oper1}
\frac{d}{d\gamma} = \frac{\partial }{\partial \gamma} + \frac{d \B u_i}{d \gamma}\cdot \frac{\partial}{\partial \B u_i}
=\frac{\partial }{\partial \gamma} + \frac{d \B u _i}{d \gamma}\cdot \frac{\partial}{\partial \B r_i} \ ,
\end{equation}
where the second equality results from from Eq.~(\ref{transformation}),
and here and below repeated subscript indices are summed over, unless
indicated otherwise.

\begin{figure}[!ht]
\hskip -1.3 cm
\includegraphics[scale = 0.35]{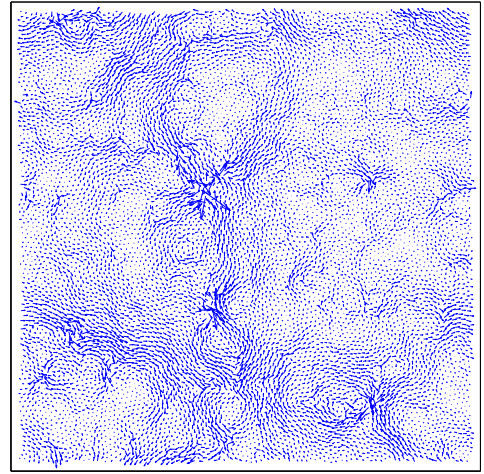}
\caption{The non-affine velocities $\B v \equiv \frac{d \B u}{d\gamma} = - \B H^{-1}\cdot \B \Xi$ for a typical
quenched realization of a two dimensional glass forming model, with $N=10000$. }
\label{nonAffVel}
\end{figure}

Using Eq.~(\ref{oper1}) to compute the stress
\begin{equation}\label{a}
\sigma = \frac{1}{V}\frac{dU}{d\gamma} = \frac{1}{V} \left[
\frac{\partial U}{\partial \gamma } + \frac{\partial U}{\partial \B r_i} \cdot \frac{d \B u_i}{d \gamma} \right]
= \frac{1}{V}\frac{\partial U}{\partial \gamma}\ ,
\end{equation}
where the second equality holds because of mechanical equilibrium $\B f_i = -\partial U/\partial {\B r}_i =0$.
The conditions of mechanical equilibrium is used further in asserting that the amorphous solid remains
in mechanical equilibrium also after straining (but before any plastic event takes place). Thus
\begin{equation}
\label{c}
\frac{d{\B f}_i}{d\gamma} = \frac{\partial {\B f}_i}{\partial \gamma }
+ \frac{\partial {\B f}_i}{\partial \B r_j} \cdot \frac{d\B u_j}{d \gamma} =0 .
\end{equation}
Using the notation for the mismatch forces
\begin{equation}
\B \Xi = -\frac{\partial \B f}{\partial \gamma} \ , \label{defXi}
\end{equation}
and the non-affine velocities
\begin{equation}
\B v = \frac{d\B u}{d \gamma} \ ,
\end{equation}
 we can rewrite Eq.~(\ref{c}) as
\begin{equation}
\label{v}
\B v = - \B H^{-1}\cdot \B \Xi\ .
\end{equation}
Here the Hessian matrix $\B H_{ij}$ is defined as  $\B H_{ij}= \frac{\partial^2 U}{\partial \B r_{i}\partial \B r_{j}}$.
A realization of the non-affine velocities is plotted in Fig.~\ref{nonAffVel}, for a two dimensional model amorphous solid.
With the definition of the non-affine velocities, the total derivative operator (\ref{oper1}) takes the form
\begin{equation}\label{oper}
\frac{d}{d\gamma} = \frac{\partial }{\partial \gamma} + \B v_i\cdot \frac{\partial}{\partial \B r_i}\ .
\end{equation}

%%%%%%%%%%%%%%%%%%%%%%%%%%%%%%%%%%%%%%%%%
%%%%%%%%%%%%%%%%%%%%%%%%%%%%%%%%%%%%%

By repeating the application of the total derivative of stress with respect to strain, the shear modulus appears
\begin{eqnarray}\label{mu}
\mu &=& \frac{1}{V}\frac{d^2U}{d\gamma^2} = \frac{1}{V}
\left[ \frac{\partial^2 U}{\partial \gamma^2} + \frac{\partial^2 U}{\partial \gamma\partial \B r_i} \cdot
\frac{d\B u_i}{d \gamma} \right] \nonumber \\
& = & \mu_B+\frac{\B v \cdot \B\Xi}{V} = \mu_B - \frac{\B \Xi \cdot \B H^{-1} \cdot \B \Xi}{V}\ ,
\end{eqnarray}
where the last step is achieved with the help of Eq.~(\ref{v}).
In Eq.~(\ref{mu}) $\mu_B$ is the affine Born contribution to the shear modulus while the second term represents
the {\bf reversible} non-affine contribution to the shear modulus due to the motion of the atoms in the amorphous solid finding new
equilibrium sites after being strained by an affine transformation. We should stress that until now the strain remains reversible in the sense that reducing
$\gamma$ back to zero (or to $\gamma_0$ if that was the reference state) returns all the coordinates $\B r_i$ to their reference
value.

Examining the expression for the shear modulus, Eq.~(\ref{mu}), we note that it contains a contraction
of the inverse of the Hessian matrix $\B H^{-1}$. The existence of eigenmodes associated
with vanishing small eigenvalues of the Hessian $\B H$ will result in a singularity in $\B H^{-1}$, which
in turn may lead to singularities in the shear modulus.
We stress here that we always consider the inverse Hessian after removing the Goldstone modes.
We will show in the following that higher order elastic coefficients consist of terms containing
an increasing number of contractions of $\B H^{-1}$, which can thus lead to increasingly
higher order singularities.

There are two main physical reasons for the existence of low-lying modes in amorphous solids.
First, it is well known that with increasing system size, one finds low lying delocalized modes, for example in the form of plane waves with wave-vector of the order of $\tilde q\sim 1/L$,
with associated eigenvalues $\tilde \lambda$ which can be estimated as
\begin{equation}
\tilde \lambda = \tilde\omega^2 \approx c^2\tilde q^2 \sim L^{-2}\sim N^{-2/d} \ .
\end{equation}

In addition to the existence of low lying delocalized modes, there exist
plastic events, which are understood as a mechanical instability in amorphous systems \cite{10KLPa,99ML,04ML}.
When a plastic event occurs at some value of the external strain $\gamma=\gamma_P$, it is manifested as a saddle node bifurcation, with the lowest eigenvalue, denoted as $\lambda_P$, vanishing like $\lambda_P\propto \sqrt{\gamma_P-\gamma}$.
It was found in \cite{10KLPc} that when $N\to \infty$ then $\gamma_P\to 0$; plastic events occur at smaller and smaller
values of the strain as the system increases in size. This important finding is reiterated below in Sect.~\ref{gammap}.
This inevitable proximity of mechanical instabilities is therefore the second reason for
the existence of low-lying modes, and below we will show that it is in fact the most dangerous one.

The question we address in this paper is how the shear modulus $\mu$ and higher order elastic moduli $B_n$
scale in the thermodynamic limit $N\to \infty$, $V\to \infty$ and $N/V=\rho$ in athermal amorphous media, by
directly analyzing the effect of low-lying modes through their corresponding statistics.
The analytic estimates are then validated against numerical simulations of model glass formers,
both in two and three dimensions.
The first point to be made is that because of the amorphous nature of the quenched solid, the
various elastic moduli are random variables and are therefore characterized by their expectation values,
their fluctuations, or more fully by their probability distribution functions (pdf's). We need to know whether
these pdf's collapse to a delta function in the thermodynamic limit. If not,
are the pdf's attaining a limit function in the thermodynamic limit, or are the fluctuations divergent in this limit.
It will turn out that $\mu$ exists in the thermodynamic limit and its distribution attains a delta function.
On the other hand, the pdf of $B_2$ does not collapse and remains wide, independently of the value of $N$.
The coefficient $B_3$ and all higher coefficients will be shown to diverge; they do not exist at all in the
thermodynamic limit. We stress that these consideration are entirely orthogonal to questions raised,
say in Ref.~\cite{10SBK} where is was argued that amorphous glassy systems at {\em finite temperatures}
can always flow if we wait sufficiently long. The present analysis has nothing to do with temperature fluctuations
or with long time scales. Our findings are also fundamentally different from those conveyed in Ref.~\cite{96BS}
where it was shown that for a solid that contains a crack the series (\ref{sigofgam}) has a zero radius of convergence.
The present analysis pertains to homogenous amorphous solid and is a direct quest for the existence of its elastic
theory in athermal conditions, but in the thermodynamic limit.

The structure of this paper is as follows; in Sect.~\ref{numerics} we describe the models and numerical methods used in
this work. In Sect.~\ref{mostDiv} we derive expressions for the most singular terms of the $n$'th order
elastic coefficients. In Sect.~\ref{gammap} we remind the reader when a freshly quenched
amorphous solid is expected to suffer its first plastic event upon increased strain,
and introduce the statistics of `plastic modes'. Sect.~\ref{planeWaves} is devoted to the
analysis of the effect of plane waves on the elastic coefficients in the thermodynamic limit;
the conclusion is that plane waves do not lead to singularities in the elastic coefficients of any order.
In Sect.~\ref{plasticModes} we present an analysis of the effect of localized plastic modes on the elastic coefficients.
We show that these modes do not lead to the divergence of the shear modulus in the thermodynamic limit.
However, they do lead to enormous fluctuations in $B_2$ - its pdf is predicted to attain a limit distribution
that does not narrow upon increasing the system size. In addition, we predict and exemplify that the plastic modes
lead to a divergence in the mean of $B_3$ in the thermodynamic limit, concluding
that it does not exist. All the higher order coefficients are expected to diverge
even faster. The paper is summarized and discussed in Sect.~\ref{summary}.
We present our conclusion regarding the physical meaning of the analysis, proposing that
our results explain some surprising experimental statements about metallic glasses
that were made by various groups.

\section{Model and Numerical Methods}
\label{numerics}
Below we employ a model glass-forming system with point particles of two `sizes' but of equal mass $m$
in two and three dimensions (2D and 3D, respectively), interacting via a pairwise potential of the form
\begin{equation}\label{potential}
\phi\left(\!\frac{r_{ij}}{\lambda_{ij}}\!\right) =
\left\{ \begin{array}{ccl} \!\!\varepsilon\left[\left(\frac{\lambda_{ij}}{r_{ij}}\right)^{k} +
\displaystyle{\sum_{\ell=0}^{q}}c_{2\ell} \left(\sFrac{r_{ij}}{\lambda_{ij}}\right)^{2\ell}\right]
&\! , \! & \frac{r_{ij}}{\lambda_{ij}} \le x_c \\ 0 &\! , \! & \frac{r_{ij}}{\lambda_{ij}} > x_c
\end{array} \right.,
\end{equation}
where $r_{ij}$ is the distance between particle $i$ and $j$, $\varepsilon$ is the energy scale, and $x_c$ is the
dimensionless length for which the potential will vanish continuously up to $q$ derivatives. The interaction
lengthscale $\lambda_{ij}$ between any two particles $i$ and $j$ is $\lambda_{ij} = 1.0\lambda$,
$\lambda_{ij} = 1.18\lambda$ and $\lambda_{ij} = 1.4\lambda$ for two `small' particles, one `large'
and one `small' particle and two `large' particle respectively. The coefficients $c_{2\ell}$
are given by
\begin{equation} c_{2\ell} = \frac{(-1)^{\ell+1}}{(2q-2\ell)!!(2\ell)!!}\frac{(k+2q)!!}{(k-2)!!(k+2\ell)}x_c^{-(k+2\ell)}.
\end{equation}
We chose the parameters $x_c = 1.48$, $k=10$ and $q=3$.
The unit of length $\lambda$ is set to be the interaction length scale of two small particles, and
$\varepsilon$ is the unit of energy and temperature. Accordingly, the time is measured in units of
$\tau_\star =\sqrt{m\lambda^2/\varepsilon}$.
The density is set to be $N/V = 0.85\lambda^{-2}$ for our two-dimensional systems,
and $N/V = 0.82\lambda^{-3}$ for our three-dimensional systems.
We prepared 5000 independent amorphous solids by quenching high temperature
equilibrium states with the rate of $10^{-4}\frac{\varepsilon}{\tau_\star}$.
Any residual heat was removed by a potential energy minimization.
We deformed our systems using the athermal quasi-static (AQS) scheme described in
detail in \cite{10KLPZ}. Elastic coefficients were calculated using the prescriptions
and microscopic expressions given in \cite{10KLPb}.

% Eigenvalues and eigenvectors of our
% quenched configurations were obtained using the LAPACK package \cite{lapack},
% and our implementation of the Lanczos algorithm \cite{lanczos}.

%%%%%%%%%%%%%%%%%%%%%%%%%%%%%%%%%%%%%%%%%%%%%%%%%%%%%%%%%%%%%%%%%%%%%%
\section{Most singular terms of elastic coefficients}
\label{mostDiv}
The shear modulus was calculated exactly in Eq.~(\ref{mu}). It contains two terms, the Born term and the
 consequence of the non-affine field $\B v$. The Born term is always finite, being a straightforward partial derivative
 of the energy (this remark pertains to the Born terms $\frac{1}{V}\partial^n U/\partial \gamma^n$ in all the higher order
 elastic coefficients as well). On the other hand, the second term in Eq.~(\ref{mu}) which is due to the non-affine response,
contains the contraction of $\B H^{-1}$ which can be potentially singular. The higher order elastic coefficients have
more terms that result from the non-affine response, but it is always easy to identify which is the potentially
most divergent term.
In the following section, we will consider the expressions for
the most singular terms of the $n$'th order elastic coefficients, denoted $D_n$, which are the terms containing the
{\bf largest} number of contractions with $\B H^{-1}$. It should be understood that $B_n$ contains terms of all orders
in $\B H^{-1}$, from the zeroth order (which is the Born term) all the way to the $2n-1$ order, cf.~\cite{10KLPa}. It is the latter
one which is potentially most singular, and we focus on that term in the sequel.

To derive the most singular terms $D_n$ of $B_n$, one can take consecutive total derivatives
of the most divergent part of $\mu$ with respect to the external strain.
The technical tool to do so is derived by taking the total derivative (\ref{oper}) with respect to strain of
the equation $\B H^{-1} \cdot \B H = \B I$, then one obtains the relation \cite{10KLPa}
\begin{equation}\label{invHesDer}
\frac{d \B H^{-1}}{d\gamma} = -\B H^{-1}\cdot
\frac{\partial \B H}{\partial \gamma}\cdot\B H^{-1} - \B H^{-1}\cdot( \B  v \cdot \B T) \cdot \B H^{-1}\ ,
\end{equation}
where we have introduced the third-rank tensor $\B T_{ijk} \equiv \frac{\partial ^3U}
{\partial \B r_k \partial \B r_j \partial \B r_i}$.
Since $\B v$ contains one contraction of $\B H^{-1}$, c.f. Eq.~(\ref{v}),
the second term on the RHS of Eq.~(\ref{invHesDer}) contains a total of {\bf three} contractions of
$\B H^{-1}$, as opposed to only one contraction in
the first term. So it is the second term $\B H^{-1}\cdot( \B  v \cdot \B T) \cdot \B H^{-1}$
that gives rise to the most singular terms which are considered below.

Using Eq.~(\ref{invHesDer}), we can operate on the most singular term of the shear modulus,
\begin{equation}\label{D1}
D_1 = -\frac{\B \Xi \cdot \B H^{-1}\cdot \B \Xi}{V} \ ,
\end{equation}
to obtain the most singular terms $D_n$ of the elastic coefficients $B_n$ of the next two orders:
\begin{equation}
D_2 = \frac{\B T \vdots \B v \B v \B v}{V}\ ,
\end{equation}
and
\begin{equation}\label{D3Tens}
D_3 = -3\frac{ (\B v\B v : \B T)\cdot \B H^{-1} \cdot (\B T : \B v\B v)}{V}\ .
\end{equation}
The full tensorial form of $D_2$ and $D_3$ was worked out in \cite{10KLPa}, and one can check directly there
that Eq.~(\ref{D3Tens}) indeed presents their most singular terms.

We denote the set of $Nd$ eigenvalues of the Hessian by
$\{\lambda_k\}_{k=1}^{Nd}$ and their associated real eigenvectors by $\{\B \Psi^{(k)}\}_{k=1}^{Nd}$.
For concreteness, we denote delocalized plane wave modes as $\tilde{\B \Psi}$, and their
associated eigenvalues as $\tilde{\lambda}$. Similarly we will denote localized plastic modes
as $\hat{\B \Psi}$, and their associated eigenvalues as $\hat{\lambda}$.
We next introduce the contractions $a_k$ and $b_{k\ell m}$:
\begin{equation}\label{overlapsDef}
a_k \equiv \B\Xi\cdot \B \Psi^{(k)}\ , \quad
b_{k\ell m} \equiv \B T\,\vdots\, \B\Psi^{(k)}\B\Psi^{(\ell)}\B\Psi^{(m)}\ .
\end{equation}
Using the normal mode decomposition of the inverse Hessian,
$\B H^{-1} = \sum_k \frac{\B \Psi^{(k)} \B \Psi^{(k)}}{\lambda_k}$,
we write the most singular terms employing the contractions $a_k$ and $b_{k\ell m}$ as
\begin{equation}\label{moses1}
D_1 = -\frac{1}{V}\sum_{k}\frac{a_k^2}{\lambda_k}\ ,\quad
D_2 = -\frac{1}{V}\sum_{k\ell m}\frac{a_ka_\ell a_m b_{k\ell m}}{\lambda_k \lambda_\ell \lambda_m}\ ,
\end{equation}
and
\begin{equation}\label{D3}
D_3 = -\frac{3}{V}\sum_{ijk\ell m}\frac{a_ia_ja_\ell a_m b_{ijk}b_{k\ell m}}
{\lambda_i\lambda_j\lambda_k \lambda_\ell \lambda_m}\ .
\end{equation}
The $n$'th most singular term can be written in a symbolic notation as
\begin{equation}\label{symbolic}
D_n = -\frac{c_n}{V} \sum_{(\ldots)}\frac{\{a\}^{n+1}\{b\}^{n-1}}{\{\lambda\}^{2n-1}}\ ,
\end{equation}
with the combinatorial factors $c_n = (2n-3)!!$, and the sum
should be understood as running over the corresponding $2n-1$ indices of the eigenvectors.
We list a number of important realizations regarding Eq.~(\ref{symbolic}):
\begin{enumerate}
\item[($i$)]{
Every index that is being summed upon appears exactly twice in the numerator
with a corresponding eigenvalue in the denominator. This is a direct result of the contraction of the
normal mode decomposition of $\B H^{-1} = \sum_{k}\frac{\B \Psi^{(k)}\B \Psi^{(k)}}{\lambda_k}$.
}
\item[($ii$)]{
The sum can consist of {\bf only} one arrangement of indices
(up to trivial permutations in identities of dummy indices). This can be understood by
examining the form of Eq.~(\ref{invHesDer}); in the various contractions, any $\B H^{-1}$
is always contracted with two tensors, never with a single one. So, for instance,
contractions of the form $\B T_{ijk}:\B H^{-1}_{jk}$ cannot appear.
Consequently, combinations of the form $b_{ikk}$ or $b_{kkk}$ cannot appear.}
\item[($iii$)]{
For odd $n$, $D_n$ is can be written as
\begin{equation}\label{oddSymbolic}
D_n = -\frac{c_n}{V} \sum_k\frac{s_k^2}{\lambda_k}\ ,
\end{equation}
where
\begin{equation}\label{sk}
s_k = \sum_{(\ldots)}\frac{\{a\}^{\frac{n+1}{2}}\{b\}^{\frac{n-1}{2}}_k}{\{\lambda\}^{n-1}}\ ,
\end{equation}
and the above sum should be understood to run over $n-1$ corresponding indices.
This is a consequence of the form $\B w \cdot \B H^{-1} \cdot \B w$ that the
most singular terms of odd order take, as seen in, e.g., Eqs.~(\ref{D1}) and~(\ref{D3Tens}).}
\end{enumerate}
The above points are consistent with the symmetry to flipping $\gamma \to -\gamma$
that require that for even $n$ $\langle D_n \rangle = 0$. One should note that although $B_n$ are defined
as the $n$'th total derivatives of stress with respect to strain, the most singular terms contain at most
a third derivative of the potential energy with respect to coordinates. This stems from the fact that the most singular
terms contain the highest number of contractions with $\B H^{-1}$; dimensional consideration then limit the order of the
energy derivatives with respect to coordinates to third order. This is the order that contributes since the first order derivative vanishes due to mechanical equilibrium, and the second order is the Hessian itself which annuls $\B H^{-1}$. Of course one
finds also mixed second order derivatives with respect to external strain and coordinates appearing through $\B \Xi$.

%%%%%%%%%%%%%%%%%%%%%%%%%%%%%%%%%%%%%%%%%%%%%%%%%%%%%%%%%%%%%%%%%%%%%%
%%%%%%%%%%%%%%%%%%%%%%%%%%%%%%%%%%%%%%%%%%%%%%%%%%%%%%%%%%%%%%%%%%%%%%
\section{distribution of plastic modes}
\label{gammap}

\begin{figure}
\includegraphics[scale = 0.50]{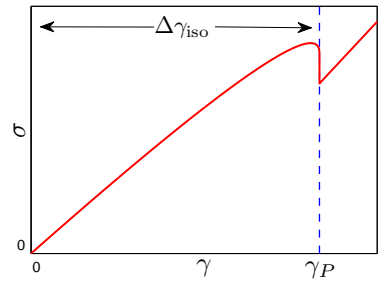}
\caption{Cartoon of a typical stress vs. strain curve of a single instance of the numerical experiment
performed in \cite{10KLPc}. Each un-deformed system was strained until the first mechanical instability
was encountered at some strain value $\gamma_P$. Statistics of $\Delta\gamma_{\rm iso} \equiv \gamma_p$
were collected for a variety of system sizes, both in two and three dimensions. }
\label{illustration}
\end{figure}

\begin{figure}
\includegraphics[scale = 0.42]{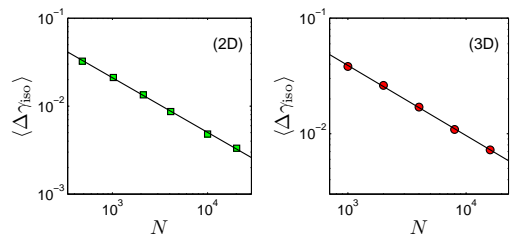}
\caption{The mean strain interval before the first mechanical instability $\langle \Delta \gamma_{\rm iso} \rangle$,
for 2D (left panel) and 3D (right panel). The continuous lines represent the scaling law Eq.~(\ref{gameq}).}
\label{gamiso}
\end{figure}

To help us in studying the issues raised in this paper, we briefly review
one of the findings of Ref.~\cite{10KLPc}, that plays a key role in the analysis of the thermodynamic limit throughout this work.
Ref.~\cite{10KLPc} employed the same generic glass-forming system in two and three dimensions as introduced
above, see \cite{10KLPc} for details. The numerical experiments performed were as follows: an ensemble of
independent initial glassy configurations quenched from the high temperature liquid was constructed.
Then the AQS scheme (see \cite{10KLPZ} for details) was utilized to strain each system up to the {\bf first}
mechanical instability occurring at some strain value $\Delta\gamma_{\rm iso}$, see Fig.~\ref{illustration} for
an illustrating cartoon. The mechanical instability is a saddle node bifurcation in which the lowest non-zero
eigenvalue $\lambda_P$ of the Hessian matrix vanishes like $\lambda_P\propto \sqrt{\gamma_P-\gamma}$. Statistics of $\Delta\gamma_{\rm iso}$ were collected for a variety of system sizes.
The {\bf first} plastic event (when a freshly quenched un-deformed system is strained) does not occur for
any infinitesimal value of $\gamma$ and careful measurement of the mean strain
interval $\langle \Delta \gamma_{\rm iso}\rangle$ that separates
the un-deformed state from the first plastic event results in a scaling law
\begin{equation}
\langle \Delta \gamma_{\rm iso} \rangle
\sim N^{\beta_{\rm iso}}\ ,\quad  \beta_{\rm iso}\approx -0.62\ . \label{gameq}
\end{equation}
In Fig.~\ref{gamiso} we show the numerical evidence that supports Eq.~(\ref{gameq}). If we accept (the reasonable but
not proven) assertion that the scaling law measured at relatively low values of $N$ continues for all $N$,
this scaling law means that the potentially unstable eigenvalue of the Hessian, $\lambda_P$, which goes through the saddle-node bifurcation at the first plastic event scales with the system size like
\begin{equation}
\lambda_P \sim \sqrt{\Delta \gamma_{\rm iso}}\sim N^{\beta_{\rm iso}/2}\ . \label{lambdaP}
\end{equation}
We should state at this point that the accuracy of measurement
cannot rule out that the second digit in the exponent is not accurate, so we need to use Eq.~(\ref{lambdaP})
with a grain of salt, as discussed below.
Based on this finding, we work out the accumulation of localized modes around $\lambda_P$.
Once a distribution of `plastic modes', $P(\hat \lambda)$, is formed in the thermodynamic limit,
we only know that when we observe its minimal value we find $\lambda_P\sim N^{\beta_{\rm iso}/2}$.
Using the Weibull theorem \cite{39Wei} we can therefore estimate that
\begin{equation}
P(\hat\lambda)\sim (N/\lambda_D)\hat\lambda^\theta K(\hat\lambda) \ , \label{PlamP}
\end{equation}
where $\theta = -\frac{2+\beta_{\rm iso}}{\beta_{\rm iso}}\approx 2.23$,
and $K(\hat\lambda)$ is some smooth function with a cutoff of the order of
a typical (Debye) cutoff eigenvalue $\lambda_D$.

%%%%%%%%%%%%%%%%%%%%%%%%%%%%%%%%%%%%%
%%%%%%%%%%%%%%%%%%%%%%%%%%%%%%%%%%%%%

%%%%%%%%%%%%%%%%%%%%%%%%%%%%%%%%%%%%%
%%%%%%%%%%%%%%%%%%%%%%%%%%%%%%%%%%%%%

%%%%%%%%%%%%%%%%%%%%%%%%%%%%%%%%%%%%%%%%%%%%%%%%%%%%%%%%%%%%%%%%%%%%%%%%%

\section{Effect of delocalized modes on elastic coefficients}
\label{planeWaves}
In this section we show that the delocalized modes in the form of plane waves do not contribute
to the divergence of the elastic coefficients of any order. A reader who is more interested in the
divergence due to the plastic modes can skip this chapter in favor of the next one.

Before we turn to the analysis, we work out the system size dependence of the contractions
Eq.~(\ref{overlapsDef}) when calculated in the context of delocalized plane-wave modes~$\tilde{\B \Psi}$.

\subsection{Scaling of the contractions $\tilde{a}$ and $\tilde{b}$ on delocalized modes}
We begin with $\tilde{a}_k \equiv \B\Xi\cdot \tilde{\B \Psi}^{(k)}$:
\begin{equation*}
\tilde{a}_k = \B\Xi \cdot \tilde{\B \Psi}^{(k)}\! = -\tilde{\B \Psi}^{(k)}\!\cdot \frac{\partial \B f}{\partial \gamma} =
-\sum_{i,j\ne i}\tilde{\B \Psi}^{(k)}_i\! \cdot \frac{\partial \B f_{ij}}{\partial \gamma} \ ,
\end{equation*}
where $\B f_{ij}$ is the force exerted by the $j$th particle on the $i$th particle. Using the symmetry of the
binary force we can rewrite
\begin{eqnarray}
\tilde{a}_k&=&-\frac{1}{2} \sum_{i,j\ne i}\tilde{\B \Psi}^{(k)}_i
\cdot\frac{\partial}{\partial \gamma} \left(\B f_{ij}-\B f_{ji}\right) \nonumber \\
&=&\frac{1}{2} \sum_{i,j\ne i}\left(\tilde{\B \Psi}^{(k)}_j - \tilde{\B \Psi}^{(k)}_i\right )
\cdot \frac{\partial \B f_{ij}}{\partial \gamma} \nonumber \\
&=&\frac{1}{2} \sum_{i,j\ne i}\tilde{\B \Psi}^{(k)}_{ij} \cdot
\frac{\partial \B f_{ij}}{\partial \gamma}\ , \label{xiOverlap}
\end{eqnarray}
where we have introduced the notation $\tilde{\B \Psi}_{ij}^{(k)} \!\!\equiv\!
\tilde{\B \Psi}_{j}^{(k)} \!\! - \tilde{\B \Psi}_{i}^{(k)}$.
For long wavelength modes $\tilde{\B \Psi}^{(k)}_{i}\sim \frac{1}{\sqrt{N}}\sin (\B q_k\cdot \B r_i)$
and we expect that the maximal contribution to the difference $\tilde{\B \Psi}_{ij}^{(k)}$ will follow the scaling
\begin{equation}\label{moses5}
\tilde{\B \Psi}_{ij} \sim \frac{q_k}{\sqrt{N}} \sim \sqrt{\frac{\tilde \lambda_k}{N}}\ .
\end{equation}
Since $\frac{\partial \B f_{ij}}{\partial \gamma}$
can be positive or negative, the sum of $N$ terms in Eq.~(\ref{xiOverlap}) cancels the $\sqrt{N}$ of the normalization
of the eigenmode and accordingly
\begin{equation}\label{aScaling}
\tilde{a}_k \sim q_k \sim \sqrt{\tilde \lambda_k}\ .
\end{equation}

We turn now to the contraction of delocalized plane-wave modes on the third-rank tensor $\tilde{b}_{k\ell m} \equiv
\B T_{k\ell m} \vdots \tilde{\B \Psi}^{(k)}\tilde{\B \Psi}^{(\ell)}\tilde{\B \Psi}^{(m)}$.
To proceed we need to use the symmetry properties of the tensor that were spelled
out in \cite{10KLPa}. Including spatial component Greek indices in the superscripts, the symmetries are
\begin{eqnarray}
T_{ijk}^{\nu\eta\chi}& = & 0 \quad\quad\quad\quad\quad\quad \mbox{if} \quad  i \ne j \ne k  \nonumber \\
T_{ijj}^{\nu\eta\chi}& = &-T_{iij}^{\nu\eta\chi} \quad\quad\quad \ \,\mbox{if}\quad i \ne j  \nonumber\\
T_{iii}^{\nu\eta\chi}  & = &  -\sum_{j\ne i}T_{iij}^{\nu\eta\chi} \\
T_{iij}^{\nu\eta\chi} &  = & T_{iji}^{\nu\chi\eta} = T_{jii}^{\chi\nu\eta} \ \; \,\mbox{if}\quad  i \ne j \nonumber \\
T_{ijk}^{\nu\eta\chi} & = & T_{ijk}^{\nu\chi\eta}\quad \mbox{and all poss. perm. of } \nu\eta\chi  \nonumber
\end{eqnarray}
Using these symmetries, it is easily verified that for general vectors $\B w,\B y, \B z$:
\begin{equation}
\sum_{ijk} \B T_{ijk}\vdots \B w_i\B y_j \B z_k =
\sFrac{1}{2}\sum_{i,j\ne i}\B T_{iij}\vdots
\B w_{ij} \B y_{ij} \B z_{ij}\ ,
\end{equation}
with $\B w_{ij} \equiv \B w_j - \B w_i$. Using this relation and Eq.~(\ref{moses5}), we can estimate
\begin{equation}\label{bScaling}
\tilde{b}_{k\ell m} \sim \frac{q_k q_\ell q_m}{N^\frac{3}{2}}\times \sqrt{N}\sim
\frac{1}{N}\sqrt{\tilde{\lambda}_k \tilde{\lambda}_\ell \tilde{\lambda}_m}\ ,
\end{equation}
where the factor of $\sqrt{N}$ accounts for summing ${\cal O}(N)$ random terms
which can be positive or negative.

\subsection{Proof of no divergence}
\label{proof}
With the scaling laws (\ref{aScaling}) and (\ref{bScaling}) for the contractions $\tilde{a}$ and $\tilde{b}$
respectively in hand, we are in the position to calculate the contribution of delocalized modes on the elastic coefficients.

We begin with the case of odd $n$; we return to the scaling of $s_k$, given in
a symbolic notation in Eq.~(\ref{sk}), and consider the sums
as running over the delocalized modes $\tilde{\B \Psi}$. Since $\{\tilde{a}\}^{\frac{n+1}{2}}\{\tilde{b}\}^{\frac{n-1}{2}}_k \sim
\sqrt{\tilde{\lambda}_k}\{\tilde{\lambda}\}^{n-1}/N^{\frac{n-1}{2}}$, then $s_k$ is a sum
over ${\cal O}(N^{n-1})$ positive or negative terms of order $\sqrt{\tilde{\lambda}_k}\{\tilde{\lambda}\}^{n-1}/N^{\frac{n-1}{2}}$.
This leaves us with $s_k \sim \sqrt{\tilde{\lambda}_k}$. Next, plugging in this scaling into (\ref{oddSymbolic}), we obtain
\begin{equation}\label{oddDn}
D_n \sim -\frac{1}{V}\sum_k \frac{s_k^2}{\tilde{\lambda}_k} \sim {\cal O}(1)\ , \quad \text{odd $n$} \ ,
\end{equation}
since it is (the negative of) a sum of $N$ positive terms, each of ${\cal O}(1)$, and
the pre-factor of $1/V$ cancels the $N$ dependence altogether.
To exemplify these arguments for odd $n$, consider $D_3$ given in Eq.~(\ref{D3}); we can rewrite it as
\begin{eqnarray}
D_3 & \sim & -\frac{3}{V}\sum_k\frac{1}{\tilde{\lambda}_k}
\sum_{ij}\frac{\tilde{a}_i\tilde{a}_j\tilde{b}_{ijk}}
{\tilde{\lambda}_i\tilde{\lambda}_j}
\sum_{\ell m}\frac{\tilde{a}_\ell \tilde{a}_m \tilde{b}_{k\ell m}}
{\tilde{\lambda}_\ell \tilde{\lambda}_m} \nonumber \\
&\sim & -\frac{3}{V}\sum_k\frac{1}{\tilde{\lambda}_k}
\left(\sum_{ij}\frac{\tilde{a}_i\tilde{a}_j\tilde{b}_{ijk}}
{\tilde{\lambda}_i\tilde{\lambda}_j}\right)^2 \ . \nonumber
\end{eqnarray}
In this case $\tilde{a}_i\tilde{a}_j\tilde{b}_{ijk} \sim \tilde{\lambda}_i\tilde{\lambda}_j\sqrt{\tilde{\lambda}_k}/N$ and
\begin{equation}
s_k = \sum_{ij}\frac{\tilde{a}_i\tilde{a}_j\tilde{b}_{ijk}}
{\tilde{\lambda}_i\tilde{\lambda}_j} \sim \sqrt{\tilde{\lambda}_k}\ ,
\end{equation}
since it is a sum of ${\cal O}(N^2)$ positive or negative terms, each of order $\sqrt{\tilde{\lambda}_k}/N$.
From here, the scaling given in Eq.~(\ref{oddDn}) is immediately obtained.

We next consider the case of even $n$; starting from Eq.~(\ref{symbolic}), we again consider sums over
the delocalized modes $\tilde{\B \Psi}$.  In the even $n$ case we consider the numerator
$\{\tilde{a}\}^{n+1}\{\tilde{b}\}^{n-1} \sim \{\tilde{\lambda}\}^{2n-1}/N^{n-1}$, then
\begin{equation}
\frac{\{\tilde{a}\}^{n+1}\{\tilde{b}\}^{n-1}}{\{\tilde{\lambda}\}^{2n-1}} \sim \frac{1}{N^{n-1}}\ .
\end{equation}
Now $D_n$ is a sum over ${\cal O}(N^{2n-1})$ positive or negative terms of
order $1/N^n$ (incorporating the $1/V$ pre-factor), hence,
for even $n$
\begin{equation}\label{evenDn}
D_n \sim \frac{1}{\sqrt{N}}\ , \quad \text{even $n$} \ .
\end{equation}
To exemplify these arguments for even $n$, consider $D_4$ which can be worked out from Eq.~(\ref{symbolic})
\begin{equation}\label{D4}
D_4 \sim -\frac{15}{V} \sum_{ijk\ell mpq}
\frac{\tilde{a}_i\tilde{a}_j\tilde{a}_k\tilde{a}_\ell\tilde{a}_m \tilde{b}_{ijp}\tilde{b}_{pkq}\tilde{b}_{q\ell m}}
{\tilde{\lambda}_i\tilde{\lambda}_j\tilde{\lambda}_k\tilde{\lambda}_\ell\tilde{\lambda}_m\tilde{\lambda}_p\tilde{\lambda}_q}\ .
\end{equation}
In this case, using again (\ref{aScaling}) and (\ref{bScaling}), we have
\[
\tilde{a}_i\tilde{a}_j\tilde{a}_k\tilde{a}_\ell\tilde{a}_m \tilde{b}_{ijp}\tilde{b}_{pkq}\tilde{b}_{q\ell m}
\sim \tilde{\lambda}_i\tilde{\lambda}_j\tilde{\lambda}_k\tilde{\lambda}_\ell\tilde{\lambda}_m\tilde{\lambda}_p\tilde{\lambda}_q/N^3\ ,
\]
with a sign which can be positive or negative. Therefore,
the sum in (\ref{D4}) runs over ${\cal O}(N^7)$ terms, each of order $1/N^3$.
This, together with the $1/V$ prefactor, results in the scaling $D_4 \sim 1/\sqrt{N}$,
as expected from Eq.~(\ref{evenDn}).

We conclude that the effect of low-lying delocalized modes on the most singular terms of the elastic
coefficients is regular; we obtained the scaling $D_n\sim {\cal O}(1)$ for odd $n$, and $D_n \sim 1/\sqrt{N}$ for
even $n$, which is consistent with the symmetries and the intensive nature
of the elastic coefficients.

\section{Effect of plastic modes on elastic coefficients}
\label{plasticModes}
The most important physical origin for low lying modes, besides the presence of
delocalized modes, is the eminent proximity of plastic failures manifested as mechanical instabilities.
In this section we analyze the effect of the resulting localized plastic modes
on the elastic coefficients of various orders. As opposed to the case of delocalized
modes (c.f.~Sect.~\ref{planeWaves}), in this case we cannot present an a-priori estimate of the system
size scaling of the contractions Eq.~(\ref{overlapsDef}) when calculated for the case of
plastic modes~$\hat{\B \Psi}$. We will need to rely on numerical information. Notwithstanding, we go as far as possible
with scaling arguments before turning to the numerical simulations.

We define the contractions for the case of localized plastic modes,
$\hat{b}_{k\ell m} \equiv \B T\,\vdots\,\hat{\B \Psi}^{(k)}\hat{\B \Psi}^{(\ell)}\hat{\B \Psi}^{(m)}$ and
$\hat{a}_k \equiv \B \Xi\cdot \hat{\B \Psi}^{(k)}$.
Since we expect the plastic modes to be highly localized, the triple contractions $\hat{b}_{k\ell m}$
can be taken to be represented approximately by delta-functions $\hat{b}_{k\ell m}\sim \delta_{k \ell}\delta_{\ell m}$.
With this approximation (whose consequence we test below against our numerics) we conclude
that all the sums appearing in Eq.~(\ref{symbolic}) (when considered to run
over the plastic modes $\hat{\B\Psi}$), are dominated by the diagonal terms,
\begin{equation}\label{diag}
D_n \sim -\frac{1}{V}\sum_k\frac{\hat{a}_k^{n+1}\hat b_k^{n-1}}
{\hat\lambda_k^{2n-1}}\ .
\end{equation}

It has been shown \cite{10KLPa,04ML}
that close to a mechanical instability, the eigenvalue $\lambda_P$ associated with the
eigenmode which is becoming unstable follows the equation of motion
\begin{equation}
\frac{d \lambda_P^{-1}}{d\gamma} \sim -\hat{a}_P\hat{b}_P\lambda_P^{-3} + {\cal O}(\lambda_P^{-1})\ , \quad \hat{b}_P\equiv \hat b_{kkk} ~\text{for}~k=P\  .
\end{equation}
This equation is analogous to Eq.~(\ref{invHesDer}), where it is given in terms of $\B H^{-1}$.
The analogous equation for $\hat\lambda^{-1}_k$ is more involved, but once we apply the above assumption that
$\hat{b}_{k\ell m}\sim \delta_{k \ell}\delta_{\ell m}$ it simplifies to an isomorphic equation 
\begin{equation}\label{eom}
\frac{d \hat \lambda_k^{-1}}{d\gamma} \sim -\hat a_k\hat b_k\hat \lambda_k^{-3} + {\cal O}(\hat \lambda_k^{-1})\ .
\end{equation}
A consequence of this result is that the product $\hat a_k\hat b_k$ has to remain independent of the system size.
We see this from Eq.~(\ref{lambdaP}) which implies that the LHS of Eq.~(\ref{eom}) is of the order of $\lambda^{-3}_k$.

The assumption that $\hat a_k\hat b_k$ is system size independent does not guarantee that $\hat a_k^{n+1}\hat b_k^{n-1}$
is also system size independent. In order to establish this, we can measure, say, $D_2$ close to a mechanical instability at $\gamma_P$. Defining $\hat{a}_P$ and $\hat{b}_P$ as
$\hat{a}_k$ and $\hat{b}_k$ for the plastic mode $\B \Psi^{(P)}$, such a measurement results in a
power law $D_2 = (\hat{a}_P^3\hat{b}_P/V)(\gamma_P - \gamma)^{-\frac{3}{2}}$ \cite{10KLPa}, allowing us to
extract the pre-factor. By taking statistics over the system size dependence of the pre-factor,
one can then deduce how $\hat{a}_k^{n+1}\hat{b}_k^{n-1}$ scales with system size.

%%%%%%%%%%%%%%%%%%%%%%%%%%%%%%%%%%%%%%%%%%%%

\begin{figure}
\includegraphics[scale = 0.45]{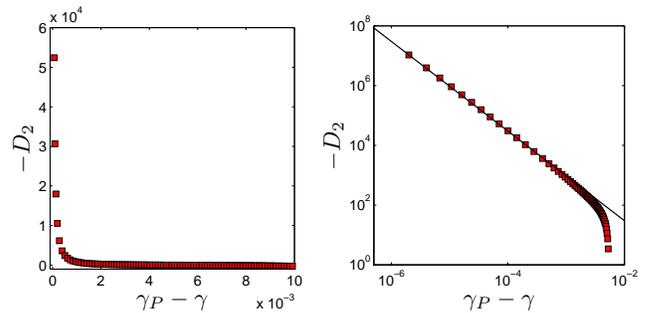}
\caption{An example of a of a single measurement of $D_2$ close to a mechanical
instability at $\gamma_P$ in 3D for $N=1000$. We extract the coefficient of the resulting power law
$D_2 \sim (\gamma_P - \gamma)^{-\frac{3}{2}}$, and average over realizations to
obtain Fig.~\ref{a2Fig}.}
\label{singleD2}
\end{figure}

\begin{figure}
\includegraphics[scale = 0.5]{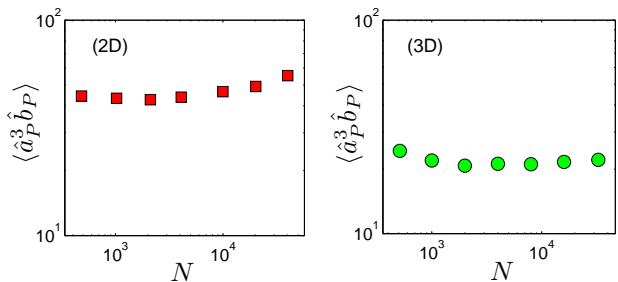}
\caption{The mean of the product of the contractions $\hat{a}_P^3\hat b_P$, as a function of system size $N$.
For the calculation below the weak observed dependence is taken as no dependence $\hat{a}_P^2\sim N^0$. }
\label{a2Fig}
\end{figure}

%%%%%%%%%%%%%%%%%%%%%%%%%%%%%%%%%%%%%%%%%%%%

We now turn to the numerical simulations. For every independent system from our ensemble
of 5000 realization per system size, we strain the system using the AQS scheme,
until the first mechanical instability is encountered at $\gamma_P$. We then backtrack and measure the
most singular term $D_2$ at various distances from the instability $\gamma_P - \gamma$.
An example of a single such measurement is plotted in Fig.~\ref{singleD2}. We then extract
the pre-factor of the resulting power law, and collect statistics of this number for various system sizes,
both in two and three dimensions. The results are plotted in Fig.~\ref{a2Fig}.
We find that the dependence of $\langle \hat{a}_P^3\hat b_P \rangle$ on $N$ is very weak. Together with
the assumption that $\hat a_P \hat b_P$ is $N$ independent we conclude that to a reasonable
approximation we can take
\begin{equation}\label{zeta}
\langle \hat{a}_P^2 \rangle\sim N^0 \ ,
\end{equation}
both in two and three dimensions. We will assume in the
following that all the $\hat{a}_k^2$ have this $N$ independence, without a
systematic dependence on the index $k$, i.e. $\langle \hat{a}_k^2 \rangle \sim
\langle \hat{a}_P^2 \rangle \sim N^0$.

Having established the scaling of $\hat{a}^2$, we next consider the sum
in Eq.~(\ref{diag}); we begin with $n=1$, and
write this sum as an integral over the density of plastic modes $P(\hat\lambda)$ given in
Eq.~(\ref{PlamP}), then
\begin{equation}
\sum_k\frac{1}{\hat\lambda_k} \to N\!\!\!\!\!\!
\int\limits_{N^{-\frac{1}{1+\theta}}}\frac{\hat\lambda^\theta K(\hat\lambda) d\hat\lambda}{\hat\lambda} \ .
\end{equation}
Since $\theta \approx 2.2 > 1$, it is obvious that the low-lying plastic modes
do not lead to a singularity in the most singular term of the first order modulus $D_1$.

For an odd $n>1$ the sum in Eq.~(\ref{diag}) becomes insensitive to the alternating signs of $\hat a_k \hat b_k$;
\begin{equation}
\sum_k\frac{\hat a_k^{n+1} \hat b^{n-1}_k}{\hat\lambda_k^{2n-1}} \sim
\sum_k\frac{1}{\hat\lambda_k^{2n-1}}\ .
\end{equation}
From here
\begin{equation}\label{noSelfAve}
\sum_k\frac{1}{\hat\lambda_k^{2n-1}} \to N\!\!\!\!\!\!
\int\limits_{N^{-\frac{1}{1+\theta}}}\frac{\hat\lambda^\theta K(\hat\lambda) d\hat\lambda}{\hat\lambda^{2n-1}} \sim
N^{\frac{2n-1}{1+\theta}}\ .
\end{equation}

For an even $n$ the sum in Eq.~(\ref{diag}) consists of positive and negative terms, and whether the sum diverges
or converges depends delicately on the exponents $\theta$ and $2n-1$. Since the mean of the moduli with even $n$
vanishes, we are interested in their pdf's (which are always symmetric around zero). The issue of convergence or
divergence is dealt in detail in Appendix \ref{sum}, where we study the pdf's of the sum
\begin{equation}\label{Xdef}
X_n(N) = \sum_k\frac{\left(g_\pm^{(k)}\right)^{n-1}}{\hat\lambda_k^{2n-1}} \ ,
\end{equation}
where $g_\pm^{(k)}$ is $+1$ or -1 with equal probabilities. We show that for the value $\theta\approx 2.2$ and
$n>1$ all these pdf's tend in the thermodynamic limit to a form with power-law tails. The result is that the re-scaled pdf
\begin{equation}
f\left(\frac{X_n(N)}{N^{\frac{2n-1}{1+\theta}}}\right) \to f(y)\sim |y|^{-\frac{2n+\theta}{2n-1}}  \ , \quad n\text{~even}, |y|\gg 1, \label{Deven}
\end{equation}
meaning that once the pdf's are rescaled by $N^{\frac{2n-1}{1+\theta}}$ they attain a limit form.

We can now finally turn to estimate the system size scaling of $D_n$;
incorporating the results of Appendix~\ref{sum} with Eqs.~(\ref{diag}),(\ref{zeta}) and (\ref{noSelfAve}), we estimate
\begin{equation}
D_n \sim \frac{1}{V}N^{\frac{2n-1}{1+\theta}}
\sim N^{\frac{2n-2-\theta}{1+\theta}}\ .
\end{equation}
For $n=2$, the prediction is
\begin{equation}
|D_2| \sim N^{-0.06}\ .
\end{equation}
Note that with our numerical accuracy regarding the exponents we cannot rule out that the exponent is actually zero. We expect therefore the distributions of $D_2$ to be either independent of system size or to decay very slowly to a delta-function.
For $n=3$, we calculate
\begin{equation}
D_3 \sim -N^{0.56}\ . \label{D3scales}
\end{equation}
Consequently $\langle D_3 \rangle$ is  predicted to diverge in the thermodynamic limit,
leading to the divergence of the third order elastic coefficients $B_3$.
Needless to say, higher order coefficients will display stronger divergences, and will
not exist in the thermodynamic limit.

\begin{figure}
\includegraphics[scale = 0.43]{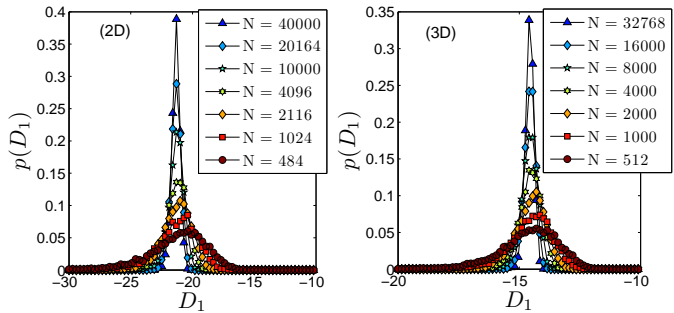}
\caption{Distributions of $D_1$ measured for each instance in our ensemble of
quenched amorphous solids. As expected, the width of these distributions decays
with increasing system size.}
\label{b1Dist}
\end{figure}

Turning again to the numerics, we measured the most singular terms $D_1$,$D_2$ and $D_3$
for each member of our ensemble of quenched amorphous solids,
both in 2D and 3D. For each member of the ensemble we computed these objects directly from their definitions
(\ref{D1}) - (\ref{D3Tens}). Due to the amorphous nature of our solid, each member of the ensemble provides
a different value to the measured quantity. The distributions of $D_1$ are displayed in Fig.~\ref{b1Dist};
as predicted, they show no anomalies, and their width decay with increasing
system size, as expected from an intensive thermodynamic variable.

%%%%%%%%%%%%%%%%%%%%%%%%%%%%%%%%%%%%%%%%
%%%%%%%%%%%%%%%%%%%%%%%%%%%%%%%%%%%%%%%%

\begin{figure}
\includegraphics[scale = 0.42]{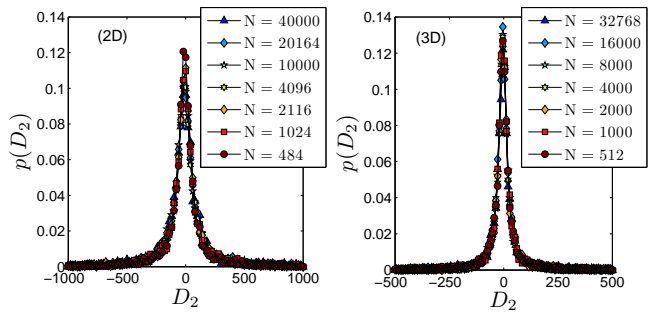}
\caption{Distributions of $D_2$ measured for each instance in our ensemble of
quenched amorphous solids. As expected, the width of these distributions does not decay
with increasing system size.}
\label{b2Dist}
\end{figure}

Next, we present the distributions of $D_2$ in Fig.~\ref{b2Dist}.
As required from isotropy, the distributions are symmetric around zero, and
$\langle D_2 \rangle = 0$. As predicted, the width of the distribution remains
$N$ independent (or decreases extremely slowly), meaning that the fluctuations over realizations of $D_2$ do not
decay appreciably with increasing the system size, in spite the intensive nature of the elastic
coefficients.

%%%%%%%%%%%%%%%%%%%%%%%%%%%%%%%%%%%%%%%%
%%%%%%%%%%%%%%%%%%%%%%%%%%%%%%%%%%%%%%%%

\begin{figure}
\hskip -0.5 cm
\includegraphics[scale = 0.45]{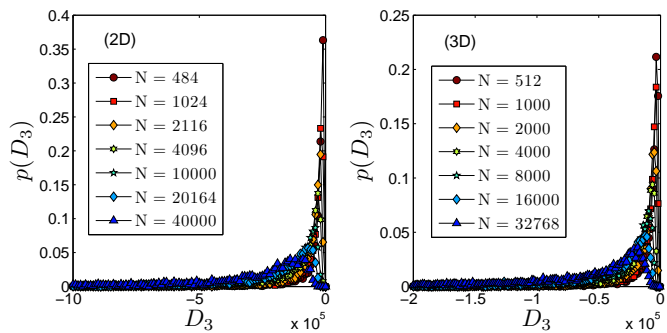}
\caption{Distributions of $D_3$ measured for each instance in our ensemble of
quenched amorphous solids.}
\label{b3Dist}
\end{figure}

\begin{figure}
\hskip -0.5 cm
\includegraphics[scale = 0.40]{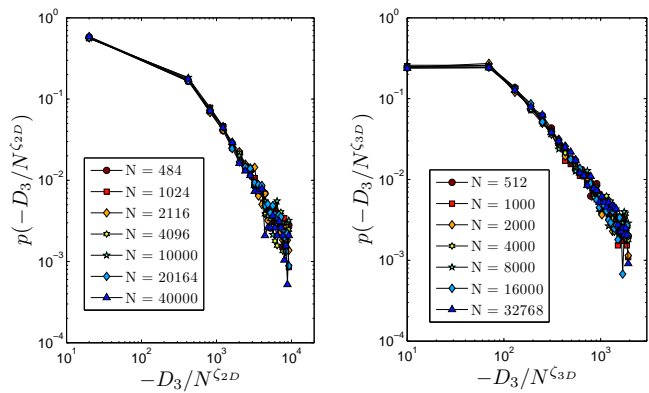}
\caption{Distributions of $D_3$ rescaled by the estimated $N$-dependence, cf.~Eq.~(\ref{D3scales}).
The data collapse is in excellent agreement with the predicted scaling behavior. We
find the best collapse with $\zeta_{2D} \approx 0.62$ for the 2D data
and $\zeta_{3D} \approx 0.56$ for the 3D data.}
\label{rescale}
\end{figure}

In Fig.~\ref{b3Dist} we display the distributions of $D_3$ measured in our
simulations. Indeed, we find that $\langle D_3 \rangle$ grows with system size, in
agreement with the estimate given here.
In Fig.~\ref{rescale} we show the distributions of $D_3$ rescaled by the expected
$N$-dependence $N^\zeta$ given by Eq.~(\ref{D3scales});
the data collapse is in very good support of the estimated
exponents. We measured approximately $\zeta_{2D} \approx 0.62$ in 2D
and $\zeta_{3D} \approx 0.56$ in 3D. Note that all the elastic coefficients from $B_3$ and higher
are expected to diverge in the thermodynamic limit even faster.

%%%%%%%%%%%%%%%%%%%%%%%%%%%%%%%%%%%%%%%%
%%%%%%%%%%%%%%%%%%%%%%%%%%%%%%%%%%%%%%%%
%%%%%%%%%%%%%%%%%%%%%%%%%%%%%%%%%%%%%%%%
%%%%%%%%%%%%%%%%%%%%%%%%%%%%%%%%%%%%%%%%

\section{Summary and Discussion}
\label{summary}
We have learned in this paper that in amorphous solids of the type discussed above the elastic theory
exists only for infinitesimal strains in the thermodynamic limit. We have found that the shear modulus exists in the thermodynamic
limits, but starting with $B_2$ the pdf's of all the nonlinear coefficients $B_n$ do not converge to a delta-function in the
thermodynamic limit. Starting with $B_3$ all the coefficient diverge with $N$. The final result can be summarized by the
prediction
\begin{equation}
B_n \sim N^{\frac{2n-2-\theta}{1+\theta}}\ .
\end{equation}
Note that the exponent $\theta$ is not expected to be universal since it depends on the quenching protocol and other
details of the system. We expect however quite generally that $\theta\ge 2$ \cite{11KLP}. Thus the shear modulus exists always, providing
us with a small linear regime where elasticity theory may remain useful.

We reiterate here that the analysis presented here
has nothing to do with time-scales. As long as our straining rate is slow enough, $\dot\gamma\ll L/c$ the formation
of an affine transformation is co-temporal with the creation of forces on the particles that result in a concurrent
non affine transformation which is the source of potential singularities in the theory. One cannot `freeze' the
non affine part of the transformation and go on straining. The `velocity' $\B v$ above is real, and represents how
the non affine transformation occurs on the same time scale that the strain is increased.

The upshot of the analysis presented above is that in amorphous solids in the thermodynamic limit it is impossible to
separate elastic from plastic responses. They are mixed together in an intimate way, defying the existence of a naive nonlinear elastic theory in the form of a Taylor expansion. It is possible that one needs to renormalize the theory in order
to obtain a well posed expansion in another variable; this possibility will be explored in a forthcoming publication.
At this point our feeling is that the results presented above explain the surprisingly high degree of plasticity that was
found in recent experiments on metallic glasses \cite{10DICAE}, and of the fact that macroscopic samples of metallic glasses cannot be maintained in the elasto-plastic steady state that is found in small sample simulations \cite{10Sam,09NM}: they
appear to break as soon as they reach the yielding transition. Also, we point out to the interesting experiments
\cite{10LWHWM, 10JG} where it was shown that in nano-samples of metallic particles one {\em could}
reach the elasto-plastic steady state. While not mathematically related to the analysis shown in this paper,
we propose that all these observations are in accordance with the theoretical conclusions presented here.
It may be worthwhile to examine systematically the system size dependence of the existence of the yielding
transition to a steady elasto-plastic state in metallic glasses and other systems of the type presented here.
In particular, it should be interesting to examine the nonlinear elastic coefficients in small system, like nano-samples
of metallic glasses, and attempt to confirm the divergence of $B_3$ and of higher order nonlinear elastic coefficients.

\acknowledgements

We thank Efim Brener for useful discussions. We also thank Matthieu Wyart for pointing out the
importance of the symmetries of the mismatch forces. This work had been supported in part by
the German Israeli Foundation, the Israel Science Foundation, the Israel Ministry of Science under
the French-Israeli collaboration and by the ERC under an advanced ``ideas" grant \# 267093 - STANPAS.

\appendix
\section{System size scaling of $X(N)$}
\label{sum}
The sum $X_n(N)$ was defined in Eq.~(\ref{Xdef}); here we provide a numerical analysis of
the statistical problem, from which we will deduce the scaling of $X_n$ with the system size $N$.
For clarity, we repeat the formulation of the problem: given a probability
distribution $K(\hat \lambda)\sim \hat{\lambda}^\theta$
for small $\hat{\lambda}$, we draw ${\cal O}(N)$ numbers from this distribution, and calculate
\begin{equation}
X_{n} = \sum_k\frac{g^{(k)}_\pm}{\hat{\lambda}_k^{2n-1}}\ ,
\end{equation}
where as before $g^{(k)}_\pm = \pm1 $ with probability 1/2 for each possibility.
Notice that here we only consider the case of even integers $n$,
then $\left(g^{(k)}_\pm\right)^{n-1} = g^{(k)}_\pm$.

\begin{figure}
\hskip -0.5 cm
\includegraphics[scale = 0.45]{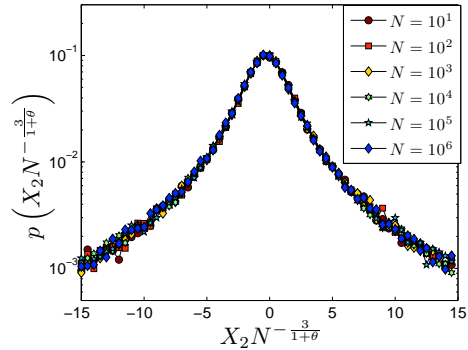}
\caption{Distributions of $X_2/N^{\frac{3}{1+\theta}}$,
calculated using $\theta = 2.2$, see text for details of statistical model.
}
\label{modelX2}
\end{figure}

\begin{figure}[!ht]
\hskip -0.5 cm
\includegraphics[scale = 0.45]{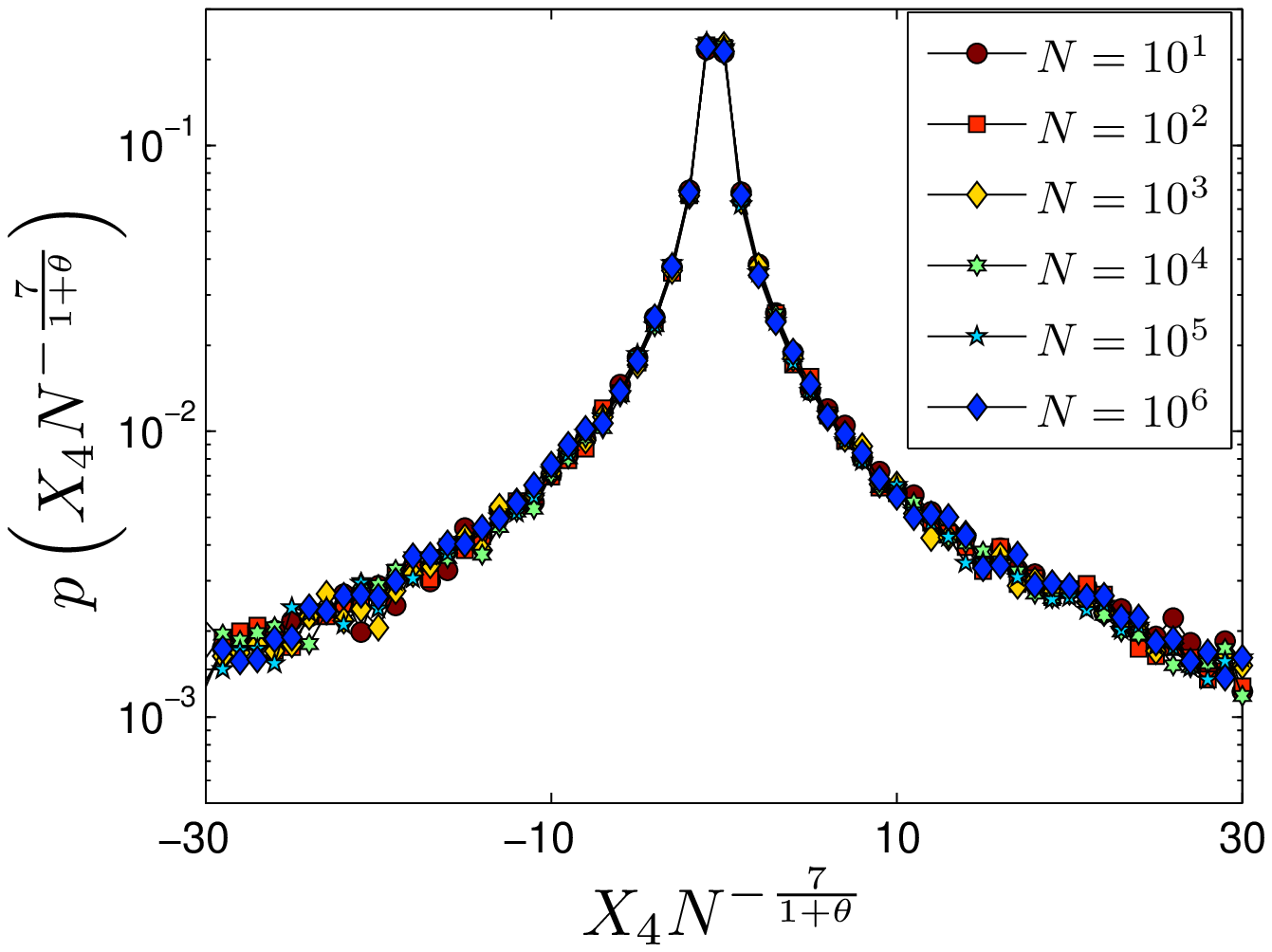}
\caption{Distributions of $X_4/N^{\frac{7}{1+\theta}}$,
calculated using $\theta = 2.2$, see text for details of statistical model. }
\label{modelX4}
\end{figure}

Following this formulation, we ran simulations of the above statistics
and extracted the pdf's of $X_n$ for $n=2$ and $n=4$,
which are the cases of interest in Sect.~\ref{plasticModes}.
The pdfs for $X_2$ and $X_4$ are plotted in semi-log scales in Figs.~\ref{modelX2} and \ref{modelX4}.
As expected from the symmetry of the factors $g^{(k)}_\pm$,
the distributions are symmetric around zero.
We confirm empirically the known result \cite{90BG} that
the pdfs of $X_n(N)$ for various $N$'s when plotted as a function of the
re-scaled argument $X_n/N^{\frac{2n-1}{1+\theta}}$, attain a limit distribution, known
as a `Levy law' \cite{90BG}. Our simulations were carried out using $\theta = 2.2$, see Sect.~\ref{gammap} and
Sect.~\ref{plasticModes} for the origin of this exponent;
we note that the extracted scaling law does $\bf not$ apply to
{\bf any} choice of exponents $\theta$ and $2n-1$,
see review by Bouchaud and Georges \cite{90BG} for further
details. According to \cite{90BG}, one expects
the limiting distributions the sums $X_{n}$ for higher orders $n>4$ to obey the {\bf same}
scaling, namely they are given by a scaling function of $X_n/N^{\frac{2n-1}{1+\theta}}$.
%%%%%%%%%%%%%%%%%%%%%%%%%%%%%%%%%%%%%%%%
%%%%%%%%%%%%%%%%%%%%%%%%%%%%%%%%%%%%%%%%

%%%%%%%%%%%%%%%%%%%%%%%%%%%%%%%%%%%%%%%%
%%%%%%%%%%%%%%%%%%%%%%%%%%%%%%%%%%%%%%%%


\begin{thebibliography}{99}
\bibitem{footnote}
The incremental strain tensor is then conventionally written as $\B \epsilon \equiv \B J^T\cdot\B J - \B I$.
\end{thebibliography}
\end{document}